\documentclass[a4paper,prl,superscriptaddress,showpacs,twocolumn]{revtex4}
\usepackage[dvips]{graphicx}
\usepackage{amssymb}
\usepackage{color}

\newcommand{\red}{\color{black}}

\begin{document}

\title{Effects of isoelectronic Ru substitution at the Fe site on the energy gaps
of optimally F-doped SmFeAsO}
\author{D. Daghero}
\affiliation{Dipartimento di Scienza Applicata e Tecnologia,
Politecnico di Torino, 10129 Torino, Italy}
\author{M. Tortello}
\affiliation{Dipartimento di Scienza Applicata e Tecnologia,
Politecnico di Torino, 10129 Torino, Italy}
\author{G.A. Ummarino}
\affiliation{Dipartimento di Scienza Applicata e Tecnologia,
Politecnico di Torino, 10129 Torino, Italy}
\author{V.A. Stepanov}
\affiliation{P.N. Lebedev Physical Institute, Russian Academy of
Sciences, 119991 Moscow, Russia}
\author{F. Bernardini}
\affiliation{Dipartimento di Fisica, Universit\`{a} di Cagliari,
09042 Monserrato (CA), Italy}
\author{M. Tropeano}
\affiliation{Dipartimento di Fisica, Universit\`{a} di Genova, via
Dodecaneso 33, 16146 Genova, Italy}
\author{M. Putti}
\affiliation{CNR-SPIN, Corso Perrone 24, 16152 Genova, Italy}
\affiliation{Dipartimento di Fisica, Universit\`{a} di Genova, via
Dodecaneso 33, 16146 Genova, Italy}

\author{R.S. Gonnelli \email{E-mail:renato.gonnelli@polito.it}}
\affiliation{Dipartimento di Scienza Applicata e Tecnologia,
Politecnico di Torino, 10129 Torino, Italy}

\begin{abstract}
We studied the effects of isoelectronic Ru substitution at the Fe
site on the energy gaps of optimally F-doped SmFeAsO by means of
point-contact Andreev reflection spectroscopy. The results show that
the $\mathrm{SmFe_{1-x}Ru_{x}AsO_{0.85}F_{0.15}}$ system keeps a
multigap character at least up to $x=0.50$, and that the gap
amplitudes $\Delta_1$ and $\Delta_2$ scale almost linearly with the
local critical temperature $T_c^A$. The gap ratios $2\Delta_i/k_B
T_c$ remain approximately constant only as long as $T_c \geq 30
\mathrm{K}$, but increase dramatically when $T_c$ decreases further.
This trend seems to be common to many Fe-based superconductors,
irrespective of their family. Based on first-principle calculations
of the bandstructure and of the density of states projected on the
different bands, we show that this trend, as well as the $T_c$
dependence of the gaps and the reduction of $T_c$ upon Ru doping,
can be explained within an effective three-band Eliashberg model as
being due to a suppression of the superfluid density at finite
temperature that, in turns, modifies the temperature dependence of
the characteristic spin-fluctuation energy.
\end{abstract}
\pacs{74.50.+r , 74.70.Dd,  74.45.+c } \maketitle


\section{Introduction}

The discovery of Fe-based superconductors (FeBS) \cite{kamihara08}
with $T_c$ as high as 55 K has shown that cuprates no longer
represent the only class of high-$T_c$ compounds. One of the reasons
of the great excitement in the scientific community and of the
impressive amount of work produced up to now is certainly that these
materials give the opportunity to study high-$T_c$ superconductivity
in different systems, in the hope to enucleate its key elements. The
parent stoichiometric compounds of most FeBS are not superconducting
(with few exceptions, like LiFeAs and LaFePO) but display a metallic
behaviour (as opposed to the Mott insulating state of cuprates) and
feature a long-range antiferromagnetic (AFM) spin-density-wave (SDW)
order. Superconductivity appears upon doping and, in some systems,
also by applying pressure; however, the order of the transition
between magnetic and superconducting phases seems not to be
universal though increasing evidences are being collected of a
region of coexistence of superconductivity and magnetism. Contrary
to cuprates (where the superconducting region in the phase diagram
is dome-shaped and the maximum $T_c$ corresponds to a well-defined
``optimal'' doping) in FeBS superconductivity sometimes appears with
$T_c$ already very close to the maximum and shows a weak doping
dependence in a broad doping range. A central feature of FeBS --
which is tightly connected to the origin of superconductivity
according to the most widely accepted theories -- is their multiband
character.  They feature indeed two or three hole pockets around the
$\Gamma$ point of the first Brillouin zone and two electron pockets
at the $M$ point (in the folded Brillouin zone)
\cite{mazin09,mazin08b}. In 1111 compounds, all the relevant Fermi
surface sheets are weakly warped cylinders parallel to the $k_z$
axis (as expected in a layered material) while a greater degree of
three-dimensionality is observed in 122 compounds. These multiple
bands and their almost perfect nesting in the parent compound
explain the AFM instability. The weakening of the nesting induced by
doping instead leads to spin fluctuations that would act as the glue
for the formation of Cooper pairs. A spin-fluctuation mediated
pairing would be mainly \emph{interband} and would favour the
opening of superconducting energy gaps of different sign on
different Fermi surface sheets, the so-called $s\pm$ symmetry
\cite{mazin08}. Though many theoretical and experimental results
support this theory \cite{chen10,inosov10} there is not, up to now,
a definitive proof of such a picture. Things are further
considerably complicated by the fact that the electronic
banstructure is very sensitive to some fine structural parameters,
like the Fe-As-Fe bond angle and more particularly the height of the
pnictogen atom ($h_{As}$) above the Fe layer. Possibly because of
this sensitivity, in many situations the gap structure of FeBS can
vary considerably within the same system, giving rise to line nodes,
point nodes, deep gap minima etc.
\cite{reid10,prozorov11,reid11,hirschfeld11}. In 1111 compounds,
$h_{As}$ has been proposed as a switch between high-$T_c$ nodeless
superconductivity and low-$T_c$ nodal superconductivity
\cite{kuroki09}.

In the effort to discriminate the effects of different parameters on
the superconducting and magnetic phases of FeBS many different
chemical substitutions have been performed. The main effect of
aliovalent substitutions is to dope the parent compound with charge,
either electrons \cite{lee09,sefat08} or holes \cite{wen08} thus
allowing to explore the phase diagram. Isovalent substitutions
\cite{tropeano10,zhigadlo11,mcguire09,lee10}, instead were tried to
modify the lattice structure, create ``chemical pressure'',
introduce disorder (acting as magnetic or non-magnetic impurities)
etc. A further degree of freedom is the site of substitution, that
can reside either in the spacing layer \cite{wen08} or in the active
one containing Fe \cite{tropeano10,zhigadlo11,mcguire09,lee10},
which is possibly directly involved in the magnetic pairing via spin
fluctuations.

Here we report on point-contact Andreev-reflection spectroscopy
(PCARS) measurements performed in
SmFe$_{1-x}$Ru$_x$AsO$_{0.85}$F$_{0.15}$ with $x$ ranging from 0 to
0.50. The considerable decrease of $T_c$ in this series of samples has
been attributed to disorder in the Fe sub-lattice \cite{tropeano10}
and/or occurrence of a short-range static magnetic order
\cite{sanna11}. PCARS results clearly indicate  the presence of a
multigap character at all the investigated levels of Ru
substitution. The superconducting gaps decrease approximately
linearly with the local critical temperature of the contact, $T_c
^A$ but, even when the latter is reduced by a factor 5 with respect
to the optimal value, they show no sign of nodes, either intrinsic
or ``accidental''. For both gaps, the $2\Delta / k_B T_c$ ratio is
rather constant down to $T_c ^A > 30 K$ but then increases
consistently below this critical temperature. Comparison with other
results in literature indicates that many different FeBS fit in this
trend, which suggests the possibility to study some properties
common to different compounds in a single samples series that allows
spanning a very wide range of critical temperatures. Thanks to
\emph{ab-initio} electronic structure calculations, the trend of the
gaps as a function of $T_c ^A$ has been reproduced within a minimal
three-band, $s \pm$ Eliashberg model. This model also takes  into
account the so-called ``feedback'' effect, i.e. the effect of the
condensate on the antiferromagnetic spin fluctuations possibly
responsible for the superconductivity in these compounds. The
evolution as a function of $T_c$ of the temperature dependence of
the condensate necessary to reproduce the experimental data looks
rather similar to that obtained from London penetration-depth
measurements performed in other FeBS, particularly in the region of
coexistence of superconductivity and magnetism. This fact suggests
that, in agreement with ref. \cite{sanna11}, proximity of
superconductivity and magnetism in these samples might be one of the
main reasons for the decrease of $T_c$ and for the observed behavior
of the energy gaps.


\section{Experimental details}
\begin{figure}
\includegraphics[width=0.9\columnwidth]{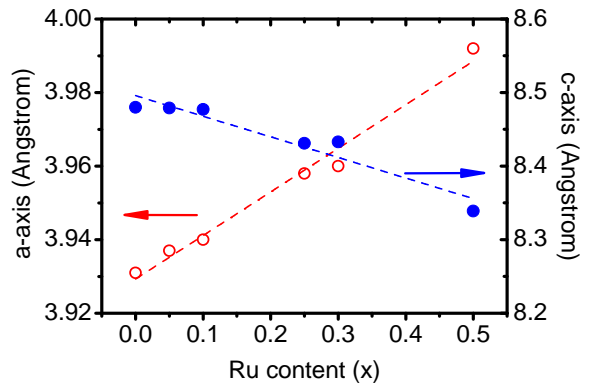}
\caption{Lattice constants for
SmFe$_{1-x}$Ru$_x$AsO$_{0.85}$F$_{0.15}$  at different Ru content
$x$. Lines are only guides to the eye.}\label{fig:1}
\end{figure}

The polycrystalline $\mathrm{SmFe_{1-x}Ru_xAsO_{0.85}F_{0.15}}$
samples were synthesized as described in Ref. \cite{tropeano10}. The
starting mixture of fine powder of SmAs and 99.9\% pure
$\mathrm{Fe_2O_3}$, $\mathrm{RuO_2}$, $\mathrm{FeF_2}$,
$\mathrm{Fe}$ and $\mathrm{Ru}$ was pressed in pellets and then put
through a two-step reaction process involving a first heating to
$450^{\circ}$ and a second heating to $1000-1075 ^{\circ}$. X-ray
diffraction analysis showed small amounts of SmOF (up to 6 \%) in
the final samples. Figure \ref{fig:1} shows the dependence of the
lattice  constants $a$ and $c$ on the Ru content $x$, indicating
that Ru substitution for Fe is effective.
Resistive critical temperatures and residual resistivities for the
samples used in this work are reported in table \ref{table samples}.
{\red The samples with $x=0.25$ and $x=0.50$ have higher $T_c$ and much
improved transport properties (namely, resistivity, magnetoresistance
and Hall mobility) than those reported in ref. \cite{tropeano10}
for the same doping contents, even though they were prepared in the same way.
The possible reason of this difference is under investigation.
In any case, these samples were particularly suited for PCARS measurements,
since the longer mean free path makes it easier to attain the spectroscopic
conditions, as explained below.}
\begin{table}
\begin{tabular}{c c c}
  \hline
  $x$ & $T_c$ (K) & $\rho_0$ (m$\Omega$ cm) \\
  \hline
  0  & 52.0  & 0.33 \\
  0.05 & 42.8 & 0.87 \\
  0.10 & 21.5 & 1.33 \\
  0.25 & 28.1 & {\red 1.20} \\
  0.30 & 13.6 & 1.69 \\
  0.50 & 13.5 & 0.70 \\
  \hline
\end{tabular}
\caption{Resistive critical temperatures and residual resistivities
(defined as in ref. \cite{tropeano10}) for
SmFe$_{1-x}$Ru$_x$AsO$_{0.85}$F$_{0.15}$ samples at different Ru
contents. {\red $T_c$ and $\rho_0$ for the samples with $x=0.25$ and $x=0.50$ are different
from those reported in ref. \cite{tropeano10}}.}\label{table samples}
\end{table}

Point contact spectroscopy is a local, surface-sensitive technique
and it is therefore necessary to avoid any surface degradation or
contamination. The samples were thus always kept in dry atmosphere,
and broken to expose a clean surface prior to point-contact
fabrication. The point contacts were made by putting a small drop of
Ag paste on that surface, as described elsewhere
\cite{gonnelli09a,daghero10}. With respect to the standard
``needle-anvil'' technique, this configuration ensures a greater
mechanical and thermal stability of the contacts and also allows the
whole mounting for point contact to be hermetically closed in the
cold head of the cryogenic insert thus avoiding any exposition to
air and moisture during the transfer from the glove box (where the
point contacts are fabricated) to the cryogenic environment.
Although the Ag drop has a diameter of at least $50 \,\mu
\mathrm{m}$, the real electric contact occurs only between some of
the Ag grains and the sample surface. The true contact is thus the
parallel of several nanoscopic junctions that can well be in the
ballistic regime {\red (i.e. have a radius smaller than the electron
mean free path). In Ref. \cite{tropeano10} a rough evaluation of
the mean free path in $\mathrm{SmFe_{1-x}Ru_xAsO_{0.85}F_{0.15}}$
gave $\ell=3- 10$ nm without any clear dependence on the Ru content.
In the cleaner samples with $x=0.25$ and $x=0.50$, the same evaluation
gives $\ell\simeq 7$ nm and $\ell \simeq 20$ nm, respectively.}
Such small values of the mean free path make the fulfillment of the ballistic
condition $a \ll \ell$ (where $a$ is the contact radius) be very difficult to achieve. For
instance, with these values of $\ell$ and the residual resistivities
taken from table \ref{table samples}, the Sharvin equation
\cite{sharvin65} would require resistances of the order of several
$\mathrm{k\Omega}$ for the contact to be ballistic.
The typical experimental resistance of the contacts is instead in
the range $10-100 \Omega$. Indeed, many of the contacts were not
spectroscopic or showed heating effects. A large number of
measurements was then necessary to achieve a relatively small number
of successful measurements. All the results reported here{\red,
except those shown in fig.\ref{fig:nonbal},} are thus referred to
the small fraction of contacts that do not show heating effects and
gave a clear Andreev-reflection signal. In these cases, the
existence of many parallel nanojunctions can be invoked to reconcile
the actual contact resistance with the requirement of ballistic
transport \cite{daghero10}. In some cases, the Sharvin condition was
fulfilled at low temperature but broke down on increasing the
temperature because of the decrease in the mean free path. In these
cases, the values of the gaps at low temperature can be taken as
meaningful anyway, though their temperature dependence and
eventually the value of the local critical temperature can be
slightly affected by the non-ideality of the contact.

\section{Results and discussion}

\subsection{Point-contact Andreev-reflection results}

\begin{figure}
\includegraphics[width=\columnwidth]{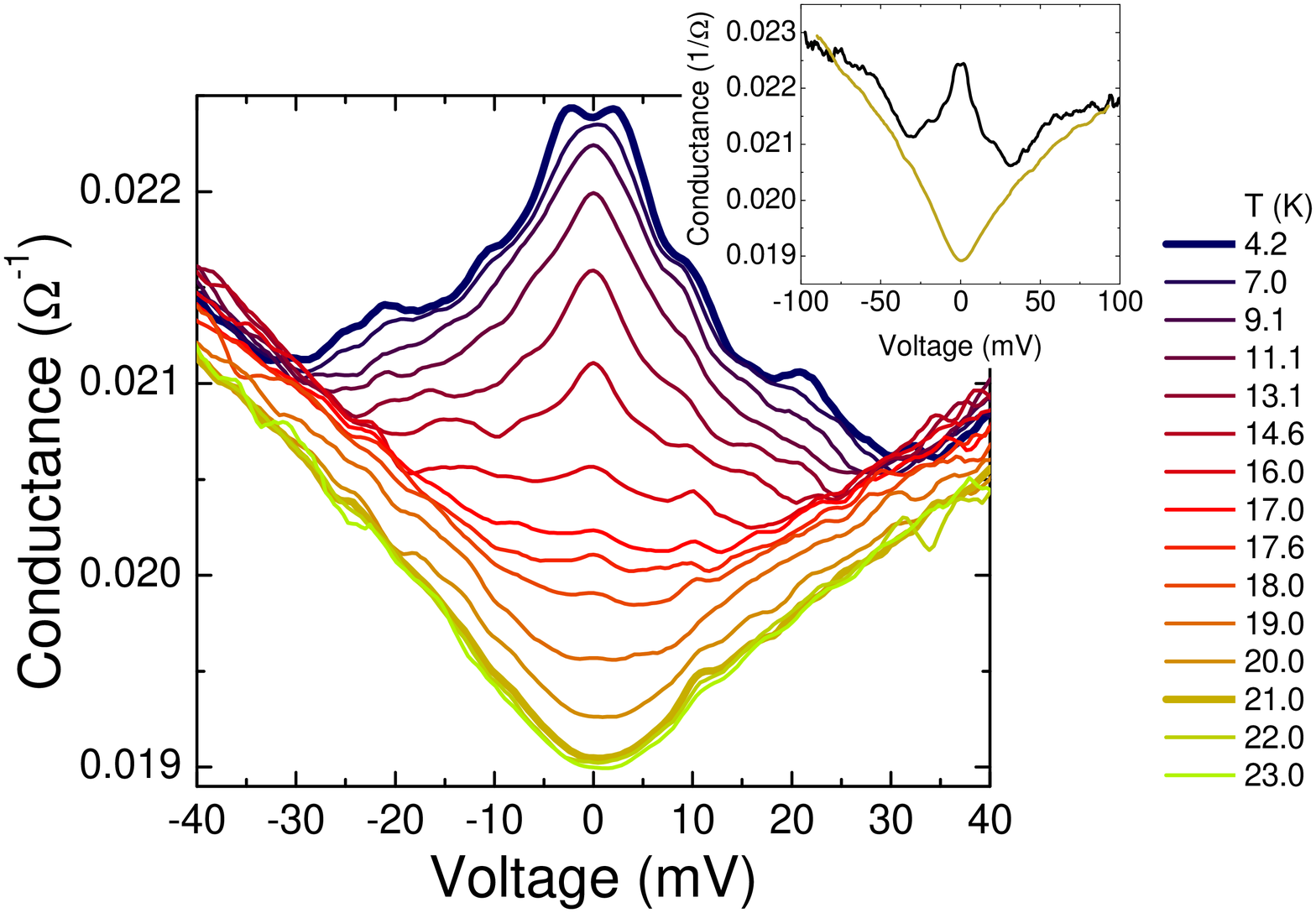}
\caption{Temperature dependence of the raw conductance curve of a
point contact on the $x=0.10$ sample. The normal state resistance is
$R_N=50 \Omega$. The inset shows the curves at 4.2 K and at 21.0 K
in an extended voltage range, to highlight the excess conductance
persisting up to about 100 mV.}\label{fig:0}
\end{figure}

Figure \ref{fig:0} reports the temperature dependence of the raw
conductance curves (obtained by numerical differentiation of the
$I-V$ characteristics) of one of the contacts that did not show any
anomaly. The curves were measured in the $x=0.10$ sample, and the
normal-state resistance of the contact is around $50\, \Omega$. The
curves show the typical features already observed in
$\mathrm{SmFeAs(O_{1-x}F_{x})}$ \cite{daghero09b}, in
$\mathrm{LaFeAs(O_{1-x}F_{x})}$ \cite{gonnelli09a} and in other 1111
compounds. In particular, they feature clear maxima related to a
presumably nodeless gap, shoulders suggestive of a second larger gap
and additional structures that, as recently shown \cite{daghero11},
can be explained as being due to the strong electron-boson coupling.
The excess conductance at high voltage, extending up to about 100 mV
(see inset) is also typical of these systems \cite{daghero11}. The
temperature at which the Andreev-reflection features disappear and
the conductance becomes equal to the normal-state one is the local
critical temperature of the contact, or Andreev critical temperature
$T_c^A$. As shown in fig.\ref{fig:0} this temperature is easy to
identify in spectroscopic contacts because it also marks the point
where conductance curves recorded at slightly different temperature
start to be superimposed to one another (here the curves at 21.0,
22.0 and 23.0 K coincide within the experimental noise).

\begin{figure}
\includegraphics[height=1\columnwidth,angle=-90]{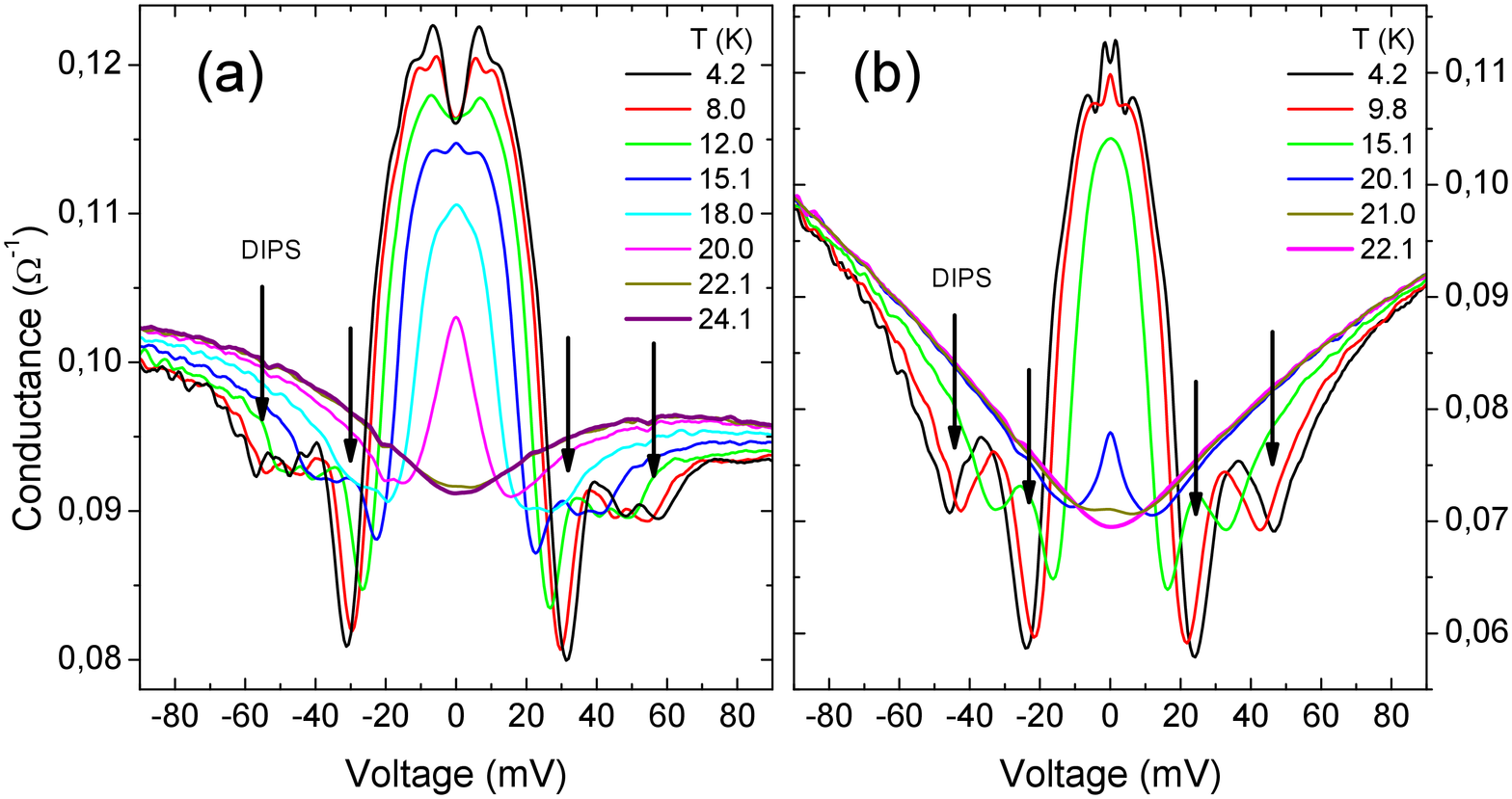}
\caption{{\red Temperature dependence of two raw conductance curves
in case of non ballistic contacts in the $x=0.1$ sample. Dip
features characteristic of non-ideal conduction through the contact
are clearly visible; the conductance minima are indicated by arrows.}}\label{fig:nonbal}
\end{figure}

{\red In contrast, Figure \ref{fig:nonbal} reports two examples of
conductance curves that show, together with an Andreev signal,
deep and wide dips  that, at low temperature, occur at energies
comparable to those of the large gap.  As shown elsewhere \cite{daghero10,sheet04}
these dips are likely to be due to the current becoming overcritical in the region of the
 contact and prevent a proper determination of the gap amplitudes. On increasing
temperature, they move toward lower voltage (due to the decrease of the critical current) causing an apparent shrinkage of the Andreev signal, finally giving rise to a sharp cusp at zero bias.}

{\red Going back to the case of ballistic contacts as in Fig.\ref{fig:0},}
the conductance at or just above $T_c^A$ can be used to
normalize all the curves at $T<T_c^A$. In principle, a conductance
curve recorded at a given temperature should be normalized to the
normal-state conductance at the \emph{same} temperature, but because
of the very high upper critical field of these materials, the latter
is not usually accessible, at least at low $T$. Using the normal
state at $T_c^A$ to normalize all the curves is thus a somehow
arbitrary choice but, as shown elsewhere \cite{daghero11}, is anyway
the one that preserves the weaker structures, i.e. those  due to the
large gap and, if present, those due to the strong electron-boson
coupling.

\begin{figure}
\includegraphics[width=0.7\columnwidth]{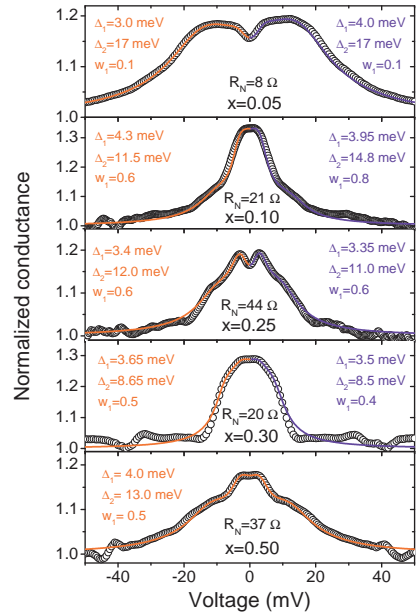}
\caption{Some examples of experimental conductance curves, after
normalization (symbols), of point contacts on samples with different
Ru content. The curves were all measured at 4.2 K. Solid blue (red)
lines represent the best fit of the right (left) side of the
experimental curves obtained within the two-band 2D generalized BTK
model \cite{kashiwaya96,daghero11}. The asymmetry decreases on
increasing the Ru vanishing completely at $x=0.50$. The fitting
parameters are indicated in the labels.}\label{fig:2}
\end{figure}

Figure \ref{fig:2} shows some examples of low-temperature,
normalized conductance curves in samples with different Ru content.
Some important points are immediately clear by looking at these
curves. First, none of them displays zero-bias peaks, and the same
happens in 100\% of the spectroscopic contacts. This points towards
the absence of line nodes even at the highest Ru contents, contrary
to what has been observed in some other FeBS away from optimal
doping \cite{reid10,prozorov11,reid11,hirschfeld11}{ \red \footnote{ {\red Moreover, none of the curves we measured showed the finite-energy peaks associated with quasiparticle
interference predicted, in some conditions,  in the nodeless $s\pm$ symmetry
\cite{golubov09}. The occurrence and the voltage position of these peaks is controlled, in the relevant theory, by a mixing parameter $\alpha$ which has not been related yet to experimental parameters. Therefore, either these peaks are not present because the conditions for their observation are not fulfilled, or they are smeared out, particularly at the $Z$ values typical of our contacts, by the broadening effects.}}}.
Second, all the curves show more or less marked double-gap features.
Third, despite the very large range of doping, the width of the
structures does not change very much (note that all the panels have
the same horizontal scale). Thus we should not expect major
variations in the gap values upon Ru doping. Fourth, the asymmetry
of the normalized conductance curves for positive/negative bias --
which is particularly strong in unsubstituted
$\mathrm{SmFeAs(O_{0.85}F_{0.15})}$ \cite{daghero09b} -- seems to be
reduced by Ru doping. As a matter of fact, it is clearly visible
even at a first glance in the case $x=0.05$, becomes discernible
only while trying to fit the data in the cases $x=0.10$ and $x=0.25$
but almost completely disappears for $x=0.30$ and $x=0.50$. The real
origin of this asymmetry, which is common to most point-contact
spectra in Fe-based superconductors, is not completely clear yet,
though it has been recently ascribed to the Seebeck effect
\cite{naidyuk10}. Preliminary Seebeck effect measurements performed
in these  samples show indeed a considerable decrease of the Seebeck
coefficient with increasing Ru content \cite{putti12}.

To extract quantitative information about the amplitude of the gaps
from the conductance curves, they must be compared with suitable
theoretical models. None of the models for single-band
superconductivity can reproduce the shape of the experimental curves
of fig.\ref{fig:2}. Instead, a two-band Blonder-Tinkham-Klapwijk
model \cite{BTK} generalized to the 2D case \cite{kashiwaya96} and
including a broadening term \cite{plecenik94} is the minimal model
that can be used in this case. {\red For each band the parameters of
the model are the energy gap $\Delta$, the broadening parameter $\Gamma$
and the barrier parameter $Z$. Then, being the total conductance the
weighed sum of the single-band conductances, the last parameter is
the weight $w_1$ of band $1$ (the weight of
band $2$ being consequently determined as $1-w_1$).\footnote{{\red The BTK model provides reliable results also in the case of parallel nano-junctions. Simulations show that the fit of the total conductance in this case gives ``effective'' parameters. The ``effective'' gap turns out to be approximately equal to the average of the gaps of the individual contacts.}}}
It is true that Sm-1111 is not two-dimensional and thus a 3D model should be used;
however, as shown elsewhere \cite{daghero11}, the latter is much
more complicated and for any practical purpose one can safely use
the 2D one (especially when, as it is the case here, the gaps are
nodeless). The lines in Fig.\ref{fig:2} represent the best fit of
the experimental data, and the labels indicate the relevant values
of the gaps and of the weight of band 1 in the point-contact
conductance, $w_1$. To account for the residual asymmetry of the
normalized curves, we chose to fit the positive- and negative-bias
side separately (blue and red lines, respectively). As previously
stated, for $x\geq 0.30$ the asymmetry is very small and the
difference between the two fits can be no longer appreciated.

\begin{figure}
\includegraphics[width=0.7\columnwidth]{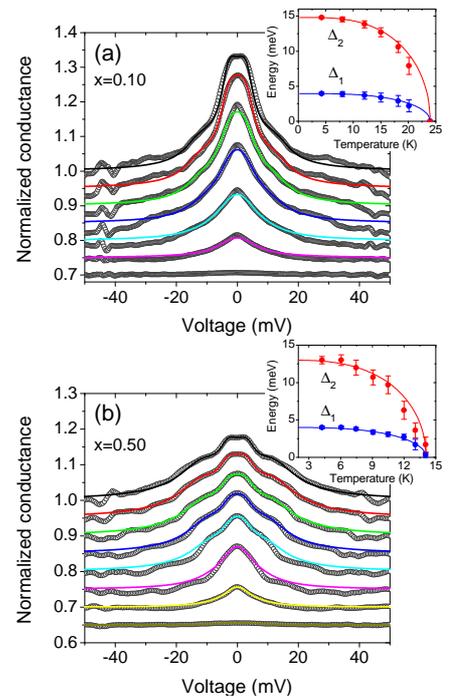}
\caption{Temperature dependence of the normalized conductance curves
(symbols) of point contacts on the samples with $x=0.10$ (a) and
$x=0.50$ (b) with the relevant two-band fit (lines). All the curves
but the top ones are vertically shifted for clarity. The insets show
the temperature dependence of the gaps $\Delta_1$ and $\Delta_2$ as
extracted from the fit. The error bars indicate the spread of gap
values obtained by different fits, when the other parameters
($\Gamma_i$, $Z_i$ and the weight $w_1$ are changed as
well).}\label{fig:3}
\end{figure}

Figure \ref{fig:3} shows two examples of how the normalized
conductance curves evolve with temperature, and the resulting
temperature dependence of the gaps extracted from the fit. The two
cases shown refer to a lightly doped sample ($x=0.10$ ) and to a
heavily doped one ($x=0.50$). The lowest-temperature curves show
clear shoulders related to the larger gap, which become less and
less discernible in the other curves (vertically offset for
clarity). In the $x=0.50$ case, a dip structure is also seen to
shift to lower energy on increasing the temperature, possibly giving
rise to the small downward deviation of the temperature dependence
of $\Delta_2$ from a BCS-like $\Delta(T)$ curve observed at high
temperature. Although there is no reason to expect the gaps to
follow a BCS-like curve, the effects of the dip do not allow us to
discuss whether this deviation is intrinsic or is an artifact due to
the small mean free path of the samples. Incidentally, on the basis
of recent calculations within a minimal three-band Eliashberg model
\cite{ummarino09} one would instead expect the gaps to be greater
than the BCS value in proximity of the critical temperature.

Let us just recall here that the fitting procedure is generally not
univocal, i.e. different sets of parameters can give almost equally
good fits. Error bars in the insets to Fig.\ref{fig:3} indicate the
spread of gap values resulting from different fits.

\begin{figure}
\includegraphics[width=0.7\columnwidth]{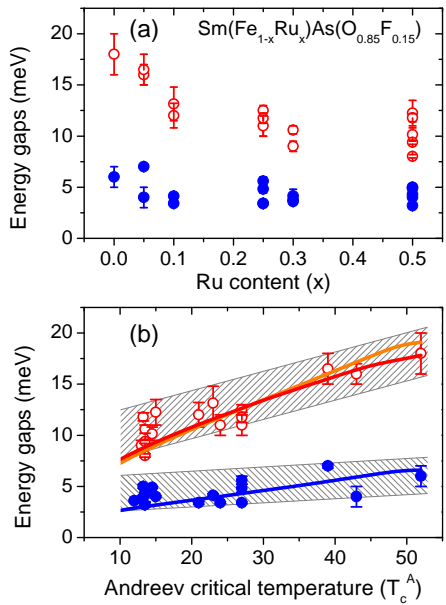}
\caption{(a) Energy gaps as extracted from the two-band fit of the
conductance curves, as a function of the Ru content $x$. (b) Energy
gaps as a function of the local critical temperature $T_c^A$ of the
various point contacts. Lines are gaps calculated within the
three-band,  $s\pm$ Eliashberg theory (see text for
details).}\label{fig:4}
\end{figure}

Figure \ref{fig:4}(a) shows the behavior of the gaps as a function
of the Ru doping $x$. The data are rather scattered but a general
trend is anyway discernible. While the small gap $\Delta_1$ does not
vary sensibly on increasing the Ru content $x$, the large gap
$\Delta_2$ shows a rapid decrease from $x=0$ to $x=0.10$ and then
remains approximately constant. This behavior is in rough
qualitative agreement with that of the bulk $T_c$ reported in table
\ref{table samples}.
{\red Since PCARS is a local probe, the
scattering of gap values at the same composition is most probably
due to slight inhomogeneities in the local doping content. As long as
$T_c$ has a strong dependence on the doping content, i.e. up to
$x=0.25$, different point contacts on the same sample can thus
provide different values of the gaps and of the local $T_c^A$, i.e. the
Andreev critical temperature.}
As a matter of fact if one plots the gaps as a function of $T_c^A$ as in
fig.\ref{fig:4}(b), a roughly linear trend of both $\Delta_1$ and
$\Delta_2$ can be appreciated despite the fluctuations in their
values. It is worth reminding that the data reported here are
already the results of a very careful selection aimed at eliminating
all the questionable results, so that these fluctuations are not due
to spurious effects that can be ascribed to non-ballistic
conduction, heating, or spreading resistance. As for the large gap,
a large uncertainty was also found in the starting compound
$\mathrm{SmFeAs(O_{1-x}F_{x})}$ \cite{daghero09b} (here represented
by the vertical error bar on the point at the maximum $T_c^A$) and
was ascribed to the residual degrees of freedom in the normalization
process, to the asymmetry of the curves and to the fact that the
features related to $\Delta_2$ are less sharp than those related to
the small gap $\Delta_1$. {\red However, in the high-doping range
($x=0.3$ and $x=0.5$), $T_c^A$ depends very little on the Ru content
and the asymmetry has almost completely disappeared; even large
differences in local composition correspond to a small difference in
$T_c^A$. Therefore, the spread of gap values accompanied by a small
spread in $T_c^A$ seems to indicate a lack of correlation between
these quantities (as observed also in MgB$_2$ with Al and Li
co-doping \cite{daghero09a}).}

\begin{figure}
\includegraphics[width=\columnwidth]{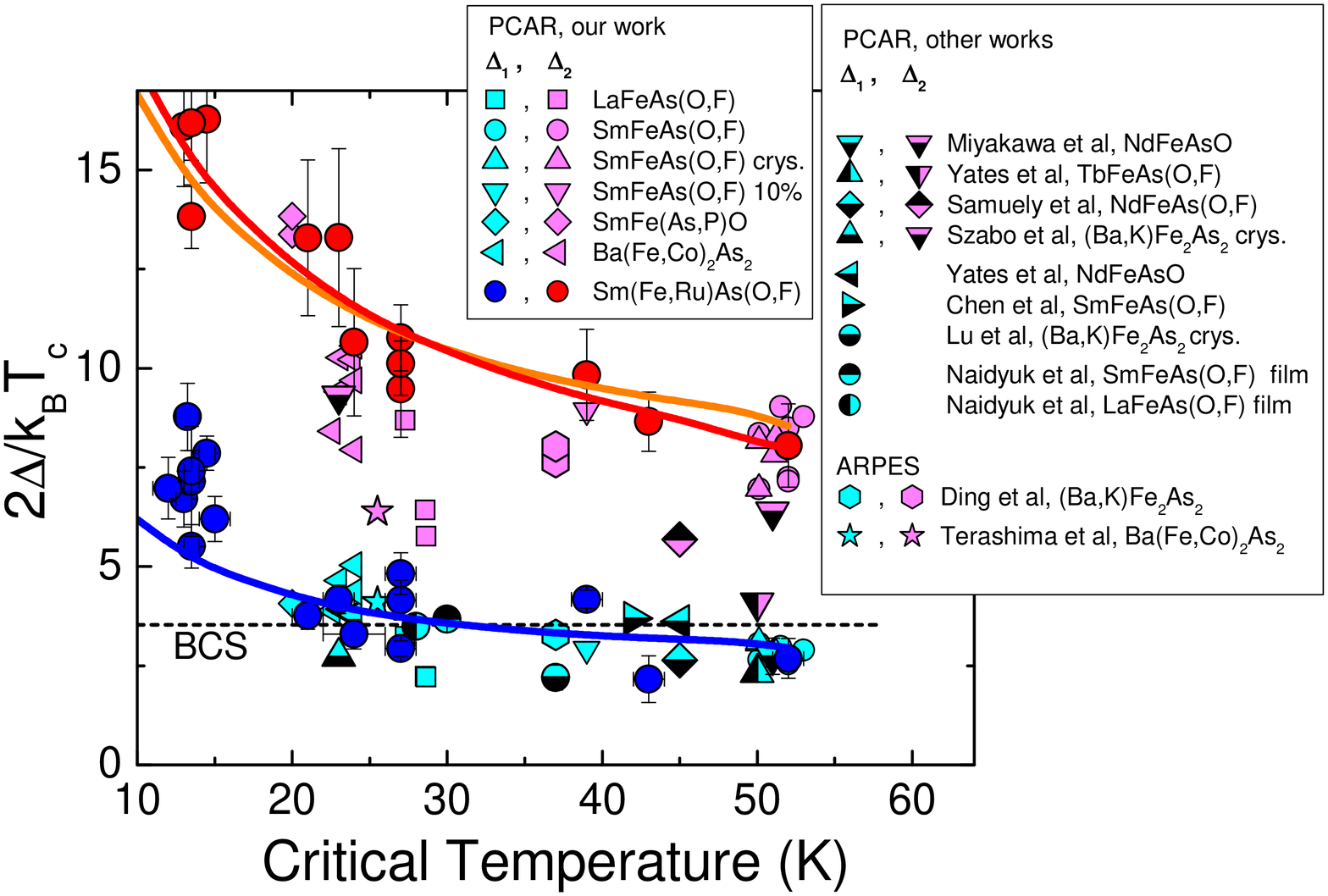}
\caption{Gap ratios $2\Delta_1/k_B T_c$ and $2\Delta_2/k_B T_c$ as a
function of $T_c$ as determined by point-contact Andreev-reflection
spectroscopy in $\mathrm{Sm(Fe_{1-x}Ru_x)As O_{0.85} F_{0.15}}$
(blue and red circles) and in various other 1111 and 122 materials.
Solid lines are  $2\Delta_i/k_B T_c$ ratios calculated within the
three-band, $s\pm$ Eliashberg theory (see text for details).
Experimental data from literature are taken from ref.
\cite{daghero11} and references therein.} \label{fig:5}
\end{figure}

Since the gaps show an overall linear trend as a function of the
local $T_c^A$, it is particularly instructive to plot the gap ratios
$2\Delta_i /k_B T_c$ as a function of $T_c^A$. This is done in
figure \ref{fig:5}, which also reports various other PCARS data in
1111 and 122 compounds. Only results showing nodeless order
parameters are shown for consistency. It is clear that the Ru
substitution in the optimally F-doped Sm-1111 allows spanning a wide
range of critical temperatures, which not only covers but also
extends the range of $T_c$ values of superconducting Fe-based
compounds measured so far by PCARS. As already shown in ref.
\cite{daghero11}, the $2\Delta_i / k_B T_c$ ratios start to increase
below $T_c \sim 30$ K. Surprisingly, PCARS data \emph{on this single
sample series}, namely SmFe$_{1-x}$Ru$_x$AsO$_{0.85}$F$_{0.15}$,
feature basically the same behavior as those obtained from many
other different nodeless FeBS of the 1111 and 122 families.

These results appear to be in contrast to what reported in ref.
\cite{inosov11}, where an opposite trend is suggested. However, even
in the aforementioned paper, several results reported for FeBS show
$2\Delta / k_B T_c$ ratios which seem to increase with decreasing
$T_c$, particularly for the large gap. A definitive answer on the
possibility of a universal trend of $2\Delta_i / k_B T_c$ vs. $T_c$
requires more experimental work, comparing results obtained with
different techniques on samples of increasingly better quality.

\subsection{Electronic structure calculations}

\begin{figure*}
\includegraphics[width=0.6\textwidth]{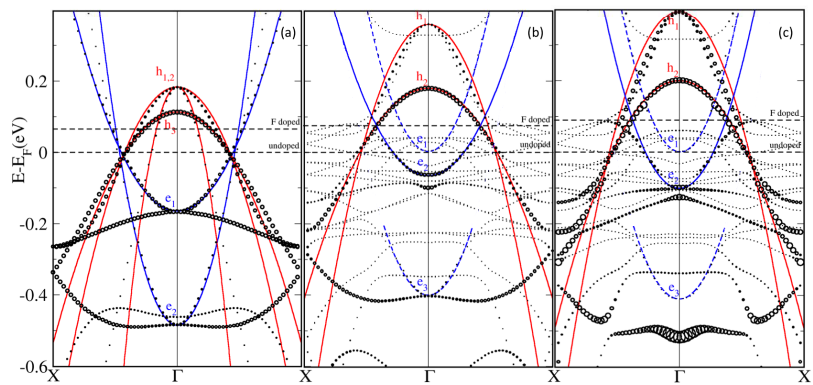}
\caption{Color online. SmRu$_x$Fe$_{1-x}$AsO fat-band structure in
the four formula unit super-cell. Dots are proportional to the Fe
{\it 3d} character of the wavefunctions. Parabolas show the results
of a fit of the Fe {\it 3d} bands within the two-dimensional
electron gas model. Upper and lower horizontal dashed lines are
E$_{\rm F}$ with and without F doping, respectively. a) Reference
band structure with La replacing Sm. b) Actual SmOFeAs band
structure including the {\it 4f} Sm bands. The two light holes bands
have similar dispersions and have been fitted with a single parabola
h$_1$. c) Same as in b) but in the case of
SmFe$_{0.75}$Ru$_{0.25}$AsO. In b) and c) the parabolic fit of band
e$_3$ is shown only up to an energy close to the one where its
hybridization with Sm {\it 4f} bands takes place.}\label{DFT}
\end{figure*}

In order to try to explain the observed PCARS data within the
Eliashberg theory, we preliminarily performed electronic structure
calculations. In particular, SmFeAsO electrons and holes density of
states (DOS) have been obtained by {\it ab initio} calculations
performed in the local density approximation to the
density-functional theory (LDA-DFT) \cite{perdew92} as implemented
in the all-electron full-potential APW and local orbitals
\cite{singh91,sjostedt00} code Wien2k \cite{wien2k}. APW has the
advantage to treat explicitly Sm {\it 4f} electrons within the
valence band yielding state-of-the-art band structure dispersion
quality. To simulate Ru substitution our calculations were performed
in a tetragonal super-cell containing four formula units ({\it Pma2}
No. 28) where 25\% Ru concentration was achieved by the substitution
of one Ru out of four Fe, retaining the bulk symmetry in the
defected cell. The conservation of symmetry allowed a reliable
comparison of doped and undoped band structures without
folding/unfolding mapping problems. Muffin tin radii of 2.3, 1.9,
2.2 and 2.0 Bohr were used for Sm, O, Fe and As, respectively.
Brillouin zone integration was performed with tetrahedrons on a
$6\times6\times4$ mesh \cite{blochl94}. Since we are interested in
the superconducting paramagnetic phase of SmFeAsO, Fe spin
polarization was not considered. The relevant band structure of this
system is given by two electron and three hole bands as it is usual
for 1111 iron-pnictides. Those are superimposed to the {\it 4f} Sm
bands at the Fermi energy (E$_{\rm F}$). To disentangle the
contribution of the Fe {\it 3d} bands we used the so-called {\it
fat-band} representation by projecting the wavefunction onto the Fe
atomic orbitals and thus obtaining the band structures shown by dots
in Fig.\ref{DFT}. The size of dots is proportional to the {\it d}
atomic character of the wavefunction on Fe atoms.

Such a procedure allows to identify a number of parabolic Fe {\it
3d} bands around E$_{\rm F}$. Since comparison with LaFeAsO is
relevant, in Fig.\ref{DFT}(a) we report the band structure obtained
for the same crystal structure of the SmFeAsO compound where Sm has
been replaced by La. In Fig.\ref{DFT}(a) we clearly see the usual
set of five parabolic Fe {\it d} bands, two electrons (e$_1$, e$_2$)
and three holes ones (h$_1$, h$_2$ and h$_3$), all of them centered
at the $\Gamma$ point since we are using a four formula unit
super-cell. This finding is in agreement with previous LDA-DFT
calculations provided we consider that SmFeAsO structure was used
here \cite{mazin08b}. As for the band structure of the actual
SmFeAsO, we find Fe {\it d} bands superimposed to Sm {\it 4f} ones.
This fact not only makes the interpretation of the band structure
less easy but introduces an hybridization effect between Sm and Fe
states. LDA bands in Fig.\ref{DFT}(b) show that Fe electron bands in
SmFeAsO are no more simply parabolic but hybridization introduces
warping out of the central part of the Brillouin zone. Hybridization
splits the electron band labeled e$_2$ in Fig.\ref{DFT}(a) in two
pieces named e$_3$ and e$_1$ in Fig.\ref{DFT}(b). In undoped SmFeAsO
e$_3$ and e$_1$ do not cross E$_{\rm F}$ being the former too low in
energy and hybridized with Sm 4f bands (see details of
Fig.\ref{DFT}(b)) and the upper too high. Therefore, undoped SmFeAsO
will have only one Fe-derived electron band e$_2$ and three hole
ones, the doubly-degenerate light-hole h$_1$ and the heavy-hole
h$_2$. The Fermi surface will be made of four nearly cylindrical
sheets, one less than LaFeAsO. Such a finding seems to be in
agreement with ARPES measurements on SmFeAsO where only one electron
band is suggested \cite{yang11}.
Since in the Eliashberg approach we are interested in the Fe {\it d}
states DOS, we need to disentangle that contribution out of the
total one that includes the Sm {\it 4f} bands. Our choice is to
model the band structure as a superposition of five parabolic bands
(including the empty e$_1$) inside the background of {\it 4f}
states. With the help of the fat-band representation, we fitted the
relevant bands along the $\Gamma$-X direction ($\Gamma$-M in the
usual two-formula unit cell) with parabolas shown as solid and
dashed lines in Fig.\ref{DFT}. Disregarding the possible warping of
the cylindrical Fermi surfaces, the DOS deriving from the above
mentioned bands was estimated from the calculated effective masses
by the free two-dimensional electron gas model
$N(E)=m^*/\pi\hbar^2$. This approximation is even more justified by
the fact that in the Eliashberg analysis reported below only ratios
between DOSs (which are much more accurate) enter in the
calculations. Then, given this assumption, only the band curvature
is relevant and therefore energy shifts with respect to LDA are not
important in the model. The only important difference relies on the
fact that the number of parabolic bands taken into consideration is
five or four. In this regard, the position of E$_{\rm F}$ is
critical since band e$_1$ is just above E$_{\rm F}$ in the undoped
compound. The superconducting phase is obtained by 15\% F doping
i.e. in SmFeAsO$_{0.85}$F$_{0.15}$. F substitution adds electrons to
the system and in the rigid band approach such effect can be coped
by the rigid shift of E$_{\rm F}$ by 0.15 electrons per formula
unit. Given the DOS $N(E)$ of the system, the shift can be simply
estimated as $0.15/N(E_{\rm F})$. Anyway, care should be taken in
defining $N(E)$, since the localized nature of Sm {\it 4f} orbitals
makes it likely that they do not receive the additional doping
charge from F. We therefore filtered out this contribution from
$N(E)$ considering only the contribution of bands e$_{1,2}$ and
h$_{1,2}$. Following this approach we get a sort of upper bound for
the E$_{\rm F}$ shift to be about 75 meV (90 meV) for
SmFeAsO$_{0.85}$F$_{0.15}$
(SmFe$_{0.75}$Ru$_{0.25}$AsO$_{0.85}$F$_{0.15}$). The Fermi levels
E$_{\rm F}$ of F-doped and undoped SmFeAsO and
SmFe$_{0.75}$Ru$_{0.25}$AsO are shown in Fig.\ref{DFT}(a)-(c). We
see that in the F-doped systems the band e$_1$ is always partially
filled justifying our assumption to include its contribution in
Eliashberg calculations. As for the effect of Ru substitution, by
comparing Fig.\ref{DFT}(b) and (c) we see that the effect is modest,
only band e$_2$ is a bit deeper and with lower effective mass as
shown in Fig.\ref{DFT}(c). Calculated DOS and plasma frequencies for
each band and for the two doping levels considered are reported in
table \ref{table}.

\begin{table}
\begin{tabular}{c c c c}
  \hline
  $x=0$ & $m_{eff}$ & DOS (st/Ha/Bohr$^2$) & $\omega_p$ (meV) \\
  \hline
  $e_1$  & 0.739  & 0.2352 & 714.04 \\
  $e_2$ & 1.508 & 0.4800 & 975.56 \\
  $h_1$ & 0.908 & 0.2890 &  1391.91 \\
  $h_2$ & 2.057 & 0.6547 & 844.86 \\
  \hline
  $x=0.25$ &  &  &  \\
  \hline
  $e_1$  & 0.668  & 0.2127 & 782.19 \\
  $e_2$ & 0.882 & 0.2809 & 1136.49 \\
  $h_1$ & 0.957 & 0.3048 &  1428.07 \\
  $h_2$ & 1.736 & 0.5526 & 864.74 \\
  \hline
\end{tabular}
\caption{Calculated effective masses, DOS and plasma frequencies for
SmFe$_{1-x}$Ru$_x$AsO$_{0.85}$F$_{0.15}$ at doping levels $x=0$ and
$x=0.25$. $h_1$ is  doubly degenerate.}\label{table}
\end{table}

\subsection{Analysis of experimental results within Eliashberg theory}

Based on the results of band structure calculations described so
far, it is possible to propose an explanation of the experimental
data by means of the simplest model that allows describing the
essential physics of the materials under study. We used the
three-band, $s\pm$ Eliashberg theory \cite{ummarino09} taking into
account the feedback effect \cite{ummarino11}. Within this model we
have two hole bands (from now on labeled as 1 and 2) and one
equivalent electron band (labeled as 3). The free parameters are
$N_i(0)$, $\lambda_{31}$, $\lambda_{32}$, $\Omega(T)$ and $\Gamma$.
$N_i(0)$ is the DOS at Fermi level, calculated above for $x=0$ and
$x=0.25$ and obtained at all the other doping levels by linear
interpolation as a function of the experimental $T_c$;
$\lambda_{31}$ and $\lambda_{32}$ are the electron-boson coupling
constants between band 3 and band 1 or 2, respectively; $\Omega(T)$
is the representative boson energy that we take as
$\Omega(T)=\Omega_0 \tanh(1.76 k \sqrt{(T_c^*/T-1)})$  where
$\Omega_0=2 T_c /5$ \cite{paglione10}; the electron-boson spectral
function has a Lorentzian shape \cite{ummarino09} with halfwidth
$\Gamma=\Omega_0/2$ \cite{inosov10}; $T_c^*$ is the feedback
critical temperature, determined by solving the Eliashberg equations
in the imaginary-axis formulation. The electron-boson coupling
matrix is

\begin{figure}
\includegraphics[width=0.9\columnwidth]{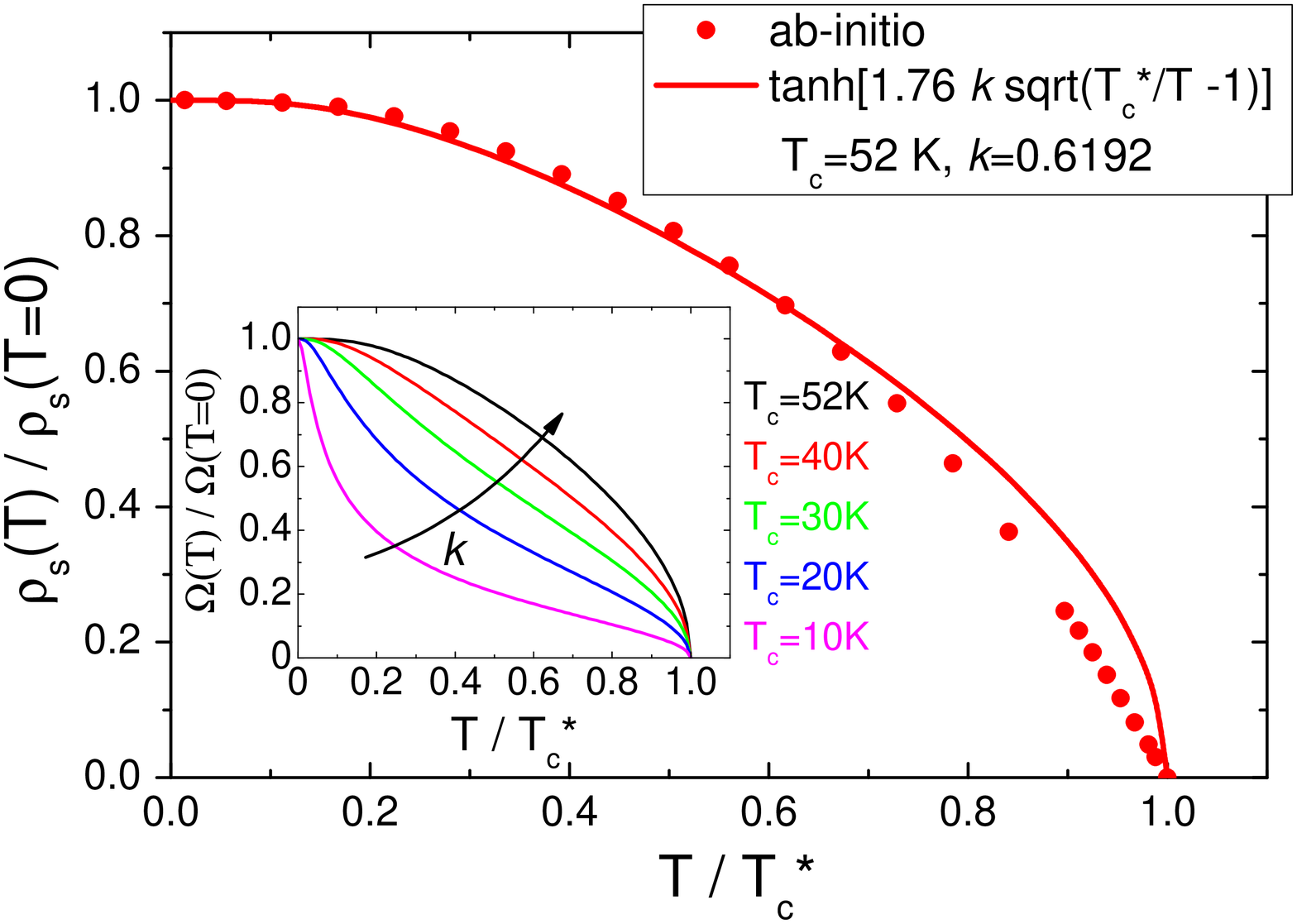}
\caption{Temperature-dependent part of the superfluid density,
$\rho_s(T)/\rho_s(T=0)$ (symbols) calculated by using the plasma
frequencies obtained from first principles calculations (see table
\ref{table}) in SmFe$_{1-x}$Ru$_x$AsO$_{0.85}$F$_{0.15}$ for $x$=0
($T_c$=52 K). The line is a fit with the formula $\tanh(1.76 k
\sqrt{(T_c^*/T-1)})$, giving $k=0.6192$. Inset:
temperature-dependent part of the superfluid density (and therefore
of the representative boson frequency \cite{ummarino11}) used in the
Eliashberg calculations at different critical temperatures (see text
for details).}\label{fig8}
\end{figure}

\vspace{-1.3mm}
\begin{displaymath}
\left(%
\begin{array}{ccc}
  0 & 0 & \lambda_{31}\nu_{1} \\
  0 & 0 & \lambda_{32}\nu_{1} \\
  \lambda_{31} & \lambda_{32} & 0 \\
\end{array}%
\right)
\end{displaymath}
%
where $\nu_{1}=N_{3}(0)/N_{1}(0)$ and $\nu_{2}=N_{3}(0)/N_{2}(0)$.
$k$ was determined at $x=0$ in the following way: first, the
superfluid density was calculated by using the plasma frequencies
obtained from first principles and reported in table \ref{table},
giving also the correct $T_c$. Then, since the temperature-dependent
part of the superfluid density corresponds to that of the
representative boson frequency \cite{ummarino11}, this curve was
fitted with $\tanh(1.76 k \sqrt{(T_c^*/T-1)})$, giving $k=0.6192$.
The values of the calculated superfluid density and the relevant fit
are shown in Fig. \ref{fig8} as symbols and line, respectively.

At this point, the only parameters that remain to be determined are
two: $\lambda_{31}$, $\lambda_{32}$. At $x=0$ they are obtained by
reproducing the two experimental gaps, but it turns out that they
also reproduce the exact experimental critical temperature. Then we
assume that the ratio $\lambda_{31}/\lambda_{32}$ is
doping-independent and that $k$ is a linear function of the
experimental critical temperature: $k(T_c)=k(T_{c, x=0})T_c/T_{c,
x=0}$. This assumption is related to the temperature dependence of
the superfluid density and is reasonable to suppose that with
increasing $x$ (and therefore decreasing of $T_c$) it decreases
\cite{prozorov11}. Now, the only free parameter left in the process
of fitting the $T_c$ dependence of the gaps is $\lambda_{31}$ which
is fixed by obtaining $T_c$ coincident with the experimental one.
The procedure is self-consistent since $\lambda_{31}$ is varied
until $T_c^*$, introduced in the formula for $\Omega(T)$, allows
reproducing the experimental $T_c$. Disorder effects have been
neglected as impurities are dominant in the intraband channel and
are thus not pair-breaking. Moreover, we also assumed that, as a
first approximation, they are also absent in the interband channel
since the two gaps are well distinct at all doping levels.

Results are shown in Fig. \ref{fig:4}(b) as solid lines: the
calculated gap values follow rather well the trend of the
experimental ones at all temperatures. The same good agreement with
the experiment can be seen also by looking at the $2\Delta_i / k_B
T_c$ ratios as a function of $T_c$ shown in Fig. \ref{fig:5} which
also reports many other results from the literature. In this regard,
it is also remarkable that it is possible, with a relatively simple
model and a small number of free parameters, to reproduce the
increase of the ratio with decreasing $T_c$. Since the strength of
the coupling increases with decreasing $T_c$, the same does, as
expected, the total electron-boson coupling constant,
$\lambda_{tot}$ which is about 3.2 at $T_c$=52 K and goes up to
almost 7.3 when $T_c$=10 K. Another interesting result that comes
out from the theoretical analysis is the temperature dependence of
the representative boson frequency (shown in the inset to Fig.
\ref{fig8} for different critical temperatures) which, as already
stated above, is equivalent to that of the superfluid density
\cite{ummarino11}. We can notice that, as $T_c$ decreases, the
superfluid density also decreases as a function of $T$ assuming,
below $T_c$=30 K, a positive curvature at intermediate temperatures.
Similar dependencies (at least in the low- and mid-$T_c$ range) have
been obtained in penetration depth measurements in Co-doped Ba-122
samples, as reported in ref. \cite{prozorov11}. In that case the
effect is less pronounced probably because the results are reported
only down to about 2$T_c$/5 while in our case $T_c$ drops to $T_c$/5
at the highest Ru doping. Moreover, it is also interesting to notice
that in Co-doped Ba-122 the temperature dependence of the superfluid
density looks slightly more depressed in samples that belong to the
region of coexistence of superconductivity and magnetism. This fact
leads us to speculate that the observed behavior of the superfluid
density in our samples might be considerably influenced by the onset
of a short-range magnetic order which competes with
superconductivity and that has been observed by $\mu$SR and
$^{75}$As NQR measurements in the same samples \cite{sanna11}.
Penetration depth measurements, as the ones in ref.
\cite{prozorov11}, would help clarifying this point as well as the
experimental determination of the temperature dependence of the
representative boson frequency, as done in refs.
\cite{inosov10,tortello10,daghero11}.

\section{Conclusions}

The isoelectronic substitution of Fe with Ru in optimally F-doped
Sm-1111 is a good way to explore a very wide range of critical
temperatures within the same Fe-based compound, in principle without
changing the total charge of the system \cite{tropeano10}. The
considerable decrease of $T_c$ induced by Ru substitution has been
ascribed to disorder in the Fe sub-lattice \cite{tropeano10} and/or
to the onset of a short-range magnetic order \cite{sanna11}. Here we
have shown that, in a wide range of $T_c$ (from 52 K down to 13.5 K,
corresponding to Ru contents ranging from 0 to 50 \%) the
$\mathrm{Sm(Fe_{1-x}Ru_x)AsO_{0.85}F_{0.15}}$ system retains its
original multi-gap character, and also the symmetry of the gaps
remains nodeless. The amplitudes of the two experimentally
detectable gaps, $\Delta_1$ and $\Delta_2$, decrease almost linearly
with $T_c$, but they remain well distinct down to the lowest $T_c$.
This suggests that the substitution-induced disorder mainly enhances
intraband scattering and does not significantly affect the interband
one. The gap ratios $2\Delta_i /k_B T_c$ strongly increase for $T_c
< 30 \mathrm{K}$ in a manner which suggests an unexpected increase
of the electron-boson coupling when $T_c$ is depressed. Very
interestingly, the trend of the gap ratios as a function of $T_c$ in
this single system is superimposed to the analogous trend obtained
by plotting the data of many Fe-based compounds of different
families \cite{daghero11}. Needless to say, this seems to point
towards a general, universal property of this class of
superconductors. By using the values of the density of states and of
the plasma frequencies calculated from first principles, we have
shown that the \emph{increase} in the gap ratios $2\Delta_i /k_B
T_c$ on decreasing $T_c$ can be reconciled with a
spin-fluctuation-mediated pairing even though the characteristic
spin-fluctuation energy has been observed to \emph{decrease}
linearly with $T_c$. The key to solving this puzzle is the feedback
effect, i.e. the effect of the condensate on the mediating boson,
which is of course only expected when the superconducting pairing
between electrons is mediated by electronic excitations
\cite{ummarino11}. An analysis carried out within an effective
three-band Eliashberg model shows indeed that the experimental
dependence of the gaps (and of the gap ratios $2\Delta_i /k_B T_c$)
on $T_c$ can be explained as being due, in particular, to a change
in the shape of the temperature dependence of the characteristic
boson energy, $\Omega_0 (T)$, with respect to the optimal-$T_c$
compound (with no Ru). Indeed, a suppression of $\Omega_0$ in the
mid-temperature range (which becomes more and more sensible on
decreasing $T_c$) is required to obtain the correct critical
temperature and the correct gap values. This finding is in very good
qualitative agreement with the experimental observation of a
depression of the superfluid density in Co-doped Ba-122 with reduced
$T_c$ \cite{prozorov11}. The fact that in the latter case this
reduction is observed in underdoped samples that fall in the region
of coexistence of magnetism and superconductivity further suggests
that, also in our samples, the depression of the condensate (that in
turns gives rise to a depression in the boson energy) at finite
temperature may be considerably influenced by the onset of a
short-range magnetic order competing with superconductivity induced
by Ru substitution, as recently observed by $\mu$SR and $^{75}$As
NQR measurements in the same set of samples \cite{sanna11}.

\section{Acknowledgements}

This work was done within the PRIN project No. 2008XWLWF9-005. FB acknowledges support from
CASPUR under the Standard HPC Grant 2012.
%


\end{document}